\RequirePackage[fleqn]{amsmath}
\documentclass[fleqn,usenatbib]{mnras}

\usepackage{txfonts}

\usepackage[T1]{fontenc}
\usepackage{ae,aecompl}

\usepackage{graphicx}	% Including figure files
\usepackage{amssymb}	% Extra maths symbols
\usepackage{wasysym}
\usepackage{soul}
\usepackage{xcolor}
\usepackage{siunitx}

\title[Halo mass function for FDM]{What is the Halo Mass Function in a Fuzzy Dark Matter Cosmology?}

\author[Kulkarni \& Ostriker]{
Mihir Kulkarni,$^{1}$\thanks{E-mail: mihir@astro.columbia.edu}
Jeremiah P. Ostriker,$^{1}$\thanks{E-mail: jpo@astro.columbia.edu}
\\
$^{1}$Columbia University, Department of Astronomy, New York, NY 10025, U.S.A.
}

\date{Accepted XXX. Received YYY; in original form ZZZ}

\pubyear{2021}

\begin{document}
\label{firstpage}
\pagerange{\pageref{firstpage}--\pageref{lastpage}}
\maketitle

\begin{abstract}
Fuzzy dark matter (FDM) or wave dark matter is an alternative theory designed to solve the small-scale problems faced by the standard cold dark matter proposal for the primary material component of the universe. It is made up of ultra-light axions having mass $\sim \SI{e-22}{eV}$ that typically have de Broglie wavelength of several kpc, alleviating some of the apparent small-scale discrepancies faced by the standard $\Lambda$CDM paradigm. In this paper, we calculate the halo mass function for the fuzzy dark matter using a sharp-k window function and compare it with one calculated using numerical simulations, finding the peak mass at roughly $\SI{e10}{M_{\odot}}$ for a particle mass of $\SI{2e-22}{eV}$. We also constrain the mass of FDM particle to be $\goa \SI{2e-22}{eV}$ using the observations of high-redshift ($z\sim 10$) lensed galaxies from CLASH survey.  
\end{abstract}

\begin{keywords}
cosmology: dark matter -- cosmology: theory -- cosmology: early universe -- galaxies: high-redshift
\end{keywords}

%%%%%%%%%%%%%%%%%%%%%%%%%%%%%%%%%%%%%%%%%%%%%%%%%%

%%%%%%%%%%%%%%%%% BODY OF PAPER %%%%%%%%%%%%%%%%%%

\section{Introduction}

The standard model of cosmology ($\Lambda$CDM) includes dark energy in the form of a cosmological constant and `cold' dark matter. This model has been immensely successful at explaining the large-scale structure of the universe, the statistics of the cosmic microwave background, and cluster abundances \citep{Bennett13}. However, recent observations have pointed out drawbacks of $\Lambda$CDM at small scales. A serious concern is the `missing satellite problem' \citep{Klypin99}. The number of satellite galaxies predicted for a Milky-way mass galaxy is greater than what we observe by an order of magnitude. This issue is sharpened by the `too big to fail' problem of galaxy formation that claims some of the predicted satellites are so massive that it is impossible for them to not have any stars \citep{Boylan11_toobigtofail}. $\Lambda$CDM also predicts a cusp in the center of the density profile of dark matter halos \citep{NFW}, whereas recent observations of dwarf galaxies suggest a flat core \citep{Burkert95, Goerdt2006}, although, it is important to note that this issue is not yet settled. In addition, the predicted dynamical friction faced by globular clusters in dwarf spheroidal galaxies is so high that the globular clusters should have spiraled and merged to the center, well before we observe them \citep{Tremaine76}.

There are typically two types of suggested solutions to these problems. The first attributes these inconsistencies to baryonic physics which is not yet very well understood. For example, the density profile of halos can be flattened to form a core when supernovae and black hole feedback redistributes matter in the galaxy \citep{Navarro96}. The missing satellite problem could be a result of baryonic physics halting the formation of galaxies or their destruction by mergers and tidal stripping. 

The other set of solutions focus on the nature of dark matter, such as warm dark matter (WDM) and fuzzy dark matter (FDM). These models suppress the small-scale structure of dark matter that results in a cut-off at the lower end of halo mass function. Warm dark matter, which is made up of less massive ($\sim \si{keV}$) particles, remains relativistic for a longer time than CDM. Its thermal velocity wipes out perturbations at small scales (free streaming). It has been pointed out that warm dark matter undergoes a `Catch-22' problem satisfying constraints simultaneously from Lyman-$\alpha$ forests and having large enough cores in dwarf galaxies \citep{Macci2012_catch22}. \cite{Macci2012_catch22} argue that having large enough cores ($\sim \SI{1}{kpc}$) in dwarf galaxies requires WDM mass to be around $\SI{0.1}{keV}$, which prevents the formation of the dwarf galaxies in the first place. 

Fuzzy dark matter which is made up of ultra-light axions with masses $\sim \SI{e-22}{eV}$ is another theory of dark matter to solve the small-scale problems \citep{Khlopov85, Hu2000_fdm}. See \cite{Hui17} for a detailed review. Their extremely low mass makes their de Broglie wavelengths typically of the order of kpc. This results in a cut-off at small-scales in the power spectrum. The fuzzy dark matter has finite quantum pressure and hence a non-zero effective sound speed given as:
\begin{equation}
c^2_{s,eff} \approx k^2/4 a^2 m_a^2,
\label{eq:cs2}
\end{equation}
where $k$ is the comoving wave number, $a$ is the scale factor and $m_a$ is the mass of axion. Thus the growth of overdensity is governed by the equation
\begin{equation}
\ddot{\delta_k} + 2 H \dot{\delta_k} + \left(\frac{c^2_{s,eff} k^2}{a^2} - 4 \pi G \rho\right) \delta_k = 0.
\label{eq:delta_growth}
\end{equation}
The solution to Eq. \ref{eq:delta_growth} describes the linear growth of perturbations with sound speed from Eq. \ref{eq:cs2}. For fuzzy dark matter, the perturbations are growing if $(c^2_{s,eff} k^2/a^2 < 4 \pi G \rho)$. The scale at which two terms are equal is called the Jeans scale,
\begin{equation}
k_J = (16 \pi G \rho a^4 m_a^2)^{1/4} = (16 \pi G \rho_0 a m_a^2)^{1/4}
\end{equation}
\citep{Hu2000_fdm}. Here, $\rho$ is the background matter density corresponding to a scale factor $a$ and $\rho_0$ is the background matter density at $z = 0$. The Jeans scale can also be re-written as
\begin{equation}
k_J = 66.5 a^{1/4} \left(\frac{\Omega_a h^2}{0.12}\right)^{1/4} \left(\frac{m}{10^{-22} eV}\right)^{1/2},
\end{equation}
as given by \cite{Marsh_review}. Here $\Omega_a$ is the ratio of the average axion matter energy density to the critical density and $h$ is defined using $H_0 = 100 h \si{km/s/Mpc}$.  A corresponding Jeans mass can be defined as: 
\begin{equation}
M_J = \frac{4 \pi}{3} \rho_0 \left(\frac{\pi}{k_J}\right)^3 \propto a^{-3/4} m_a^{-3/2}.
\end{equation}

The Jeans mass at $z = 0$ is $\sim \SI{2e7}{M_{\odot}}$. Perturbations corresponding to scales slightly larger than the Jeans scale at $z = 0$ are growing in time, however their amplitudes are highly suppressed, as they are smaller than the Jeans scale at some earlier time. An important scale that determines the relative suppression of amplitudes is the Jeans scale at matter-radiation equality, $k_{Jeq} = 9 (m_a/\SI{e-22}{eV})^{1/2} \textrm{Mpc}^{-1}$. The power spectrum for fuzzy dark matter is calculated using a redshift independent transfer function from \cite{Hu2000_fdm};
\begin{equation}
P_{FDM}(k,z) = T_F^2(k) P_{CDM}(k,z),\,  T_F(k) = \frac{\cos x^3}{1+x^8},
\label{eq:transfer function}
\end{equation}
where $x = 1.61 (m_a/\SI{e-22}{eV})^{1/18} k/k_{Jeq}$. This transfer function is determined by the Jeans scale at matter-radiation equality. As the Jeans length decreases with time, the perturbations at smaller scales continue to grow with time. With the assumption of the redshift independence of transfer function, we assume that the perturbations on all scales continue to grow in time as $\delta \propto D_+$, slightly overestimating the small scale structure.

The transfer function in Eq. \ref{eq:transfer function} is for matter-radiation equality, as it depends on the Jeans scale at that epoch. If we take the initial power spectrum at matter-radiation equality and evolve it numerically using Eq. \ref{eq:delta_growth} to $z = 0$, the shape of the power spectrum remains largely unchanged because the scales smaller than the Jeans scale at matter-radiation equality will still be highly suppressed, even if they start growing at a later epoch. This confirms the redshift independence of the transfer function.

In this paper, we calculate the halo mass function using an extended Press-Schechter formalism \citep{Press_schechter74, Bond91}. We first summarize the previous calculations of the halo mass function by \cite{Marsh_silk_14, Bozek15, Du_newpaper} that use a spherical top hat window function and a mass dependent critical density. We then point out inconsistencies in their methods and argue why a sharp-k window function works better. We compare our calculations with the halo mass function calculated using collision-less numerical simulations based on the FDM initial power spectrum \citep{Schive16_simulation}. 

Early galaxy formation can be used to constrain the properties of FDM, since the short wavelength cutoff in FDM greatly delays galaxy formation at high redshift. Thus, we also use observations of high redshift ($z \sim 10$) lensed galaxies from CLASH survey to constrain the mass of fuzzy dark matter, following a procedure similar to one used by \cite{Lensing_pacucci13} to constrain warm dark matter. 

We use cosmological parameters consistent with WMAP9 data \citep{WMAP9} ($\Omega_{m0} = 0.284$, $\Omega_{\Lambda} = 0.716$, $h = 0.696$) so as to be able to compare our results with \cite{Schive16_simulation}. 

\section{Calculations of halo mass function}
\subsection{Summary of previous calculations}

Extended Press-Schechter formalism is widely used for calculating halo mass function for dark matter using the linear power spectrum \citep{Press_schechter74, Bond91, Sheth_tormen2002}. 

Variance of perturbation amplitudes in real space smoothed over a scale $R$ is defined as follows:
\begin{equation}
S(R) = \sigma^2(R) = \int_0^{\infty} \frac{k^2 dk}{2\pi^2} P(k) W^2(kR).
\end{equation}
A spherical top hat window function is typically used, defined as
\begin{equation}
W_{TH}(k R) = \frac{3(\sin(kR) - kR \cos(kR))}{(kR)^3}.
\end{equation}

We calculate the trajectory $\delta_S$ for a point in space by starting with a large sphere of radius $R$ and decreasing the radius, and calculating the smoothed density contrast for each $R$. Use of a sharp-k window function makes increments in $\delta_S$ independent of previous steps, as $\delta(k)$ are independent Gaussian processes for different $k$. This use of the sharp-k window function makes this problem analytically solvable.  The Press-Schechter (PS) ansatz \citep{Press_schechter74} equates mass element fraction with $\delta_S > \delta_c$ with the mass fraction at time that resides in halos of mass $>M$. \cite{Bond91} removed cloud-in-cloud inconsistency in PS ansatz by equating the fraction of trajectories with first upcrossing $\delta_S = \delta_c$ at $S > S_1 = \sigma^2(M)$ with the fraction that resides in halos of mass $M<M_1$ (extended Press Schechter formalism). 

Following a few steps, we obtain the halo mass function, which is the comoving number density of halos per logarithmic mass bin, given as
\begin{equation}
\frac{dn}{d\ln M} = - \frac{\rho_0}{M} f(\sigma) \frac{d\ln \sigma}{d\ln M}. 
\end{equation}
\begin{equation}
f_{ST}(\sigma) = A \sqrt{\frac{2 a}{\pi}} \left[1+ \left(\frac{\sigma^2}{a \delta_c^2}\right)^p \right] \frac{\delta_c}{\sigma} \exp \left(\frac{a \delta_c^2}{2 \sigma^2}\right)
\end{equation}
is the fitting function for Sheth-Tormen mass function, with $A = 0.3222$, $a = 0.707$ and $p = 0.3$. The critical density $\delta_c = 1.686$ is the density contrast at collapse in linear theory for a spherical collapse model.

\begin{figure}
\includegraphics[width=\linewidth]{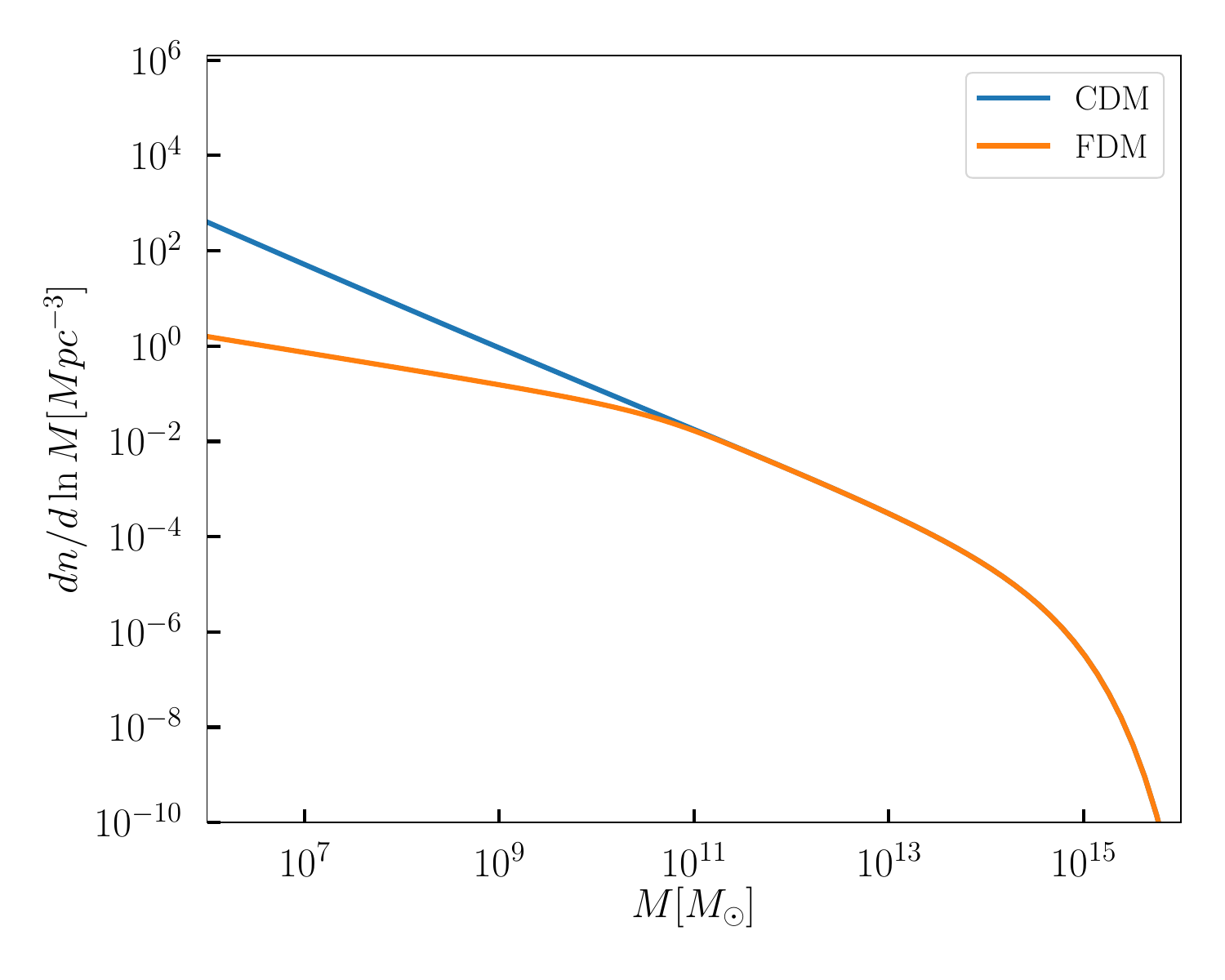}
\caption{Halo mass functions for CDM and FDM calculated with the extended Press-Schechter formalism using a top hat window function and a constant critical density for $z = 0$. We can see that the halo mass function for FDM does not have a cut-off here as expected.}
\label{fig:tophat hmfs}
\end{figure}

The halo mass function calculated using this method matches well with the numerical simulations for CDM. But if we follow the same procedure for the fuzzy dark matter, we do not obtain a cut-off in the halo mass function as expected. Figure \ref{fig:tophat hmfs} shows the halo mass functions for CDM and FDM calculated using a top-hat window function and a constant critical density. This produces halos of masses smaller than the Jeans mass for FDM, which is unphysical. The main reason for this is that the top hat window collects the highest contribution from very large scales (small-$k$ modes). For CDM, the dimensionless power spectrum $\Delta^2_k = k^3 P(k)/(2\pi^2)$ increases for increasing $k$ and hence the spherical top hat window reflects contributions from the scales of interest. FDM has a cut-off in the power spectrum and hence, while calculating $\sigma(R)$ for small $R$, the major contribution still comes from the large scales in the power spectrum, not reflecting power spectrum amplitudes at corresponding scales.

This can be understood analytically as follows:
\begin{equation}
    \frac{d\sigma^2 (R)}{d R} = \int_0^\infty \frac{k^2 dk}{2 \pi^2} P(k) \frac{d W_R^2(k)}{d R},
    \label{eq:d2sigdr}
\end{equation}

\begin{equation}
W_{TH}'(x) = \frac{3}{x^4} ((x^2 - 3)\sin x+3 x \cos x) = \frac{3}{x}\left(\frac{sin(x)}{x} - W_{TH}(x)\right),
\end{equation}

\begin{equation}
\frac{d W_{TH}^2}{d R} = \frac{6 W_{TH}(k R)}{R} \left(\frac{\sin (k R)}{k R} - W_{TH}(k R)\right),
\end{equation}
which in the limit of $R \rightarrow 0$ becomes $\propto -R$. This tells us that the derivative of the window function has a tail stretching over small $R$. Plugging this into Eq. \ref{eq:d2sigdr} gives that $\frac{d \sigma^2(R)}{d R} \propto R$ for small R. So, 
\begin{equation}
    \frac{dn}{d\ln M} = - \frac{\rho_0}{M} f(\sigma) \frac{d\ln \sigma}{d\ln M} \propto - \frac{1}{M} \frac{d\ln \sigma}{d\ln M} \propto -\frac{1}{M} R \frac{d W_{TH}^2}{d R} \propto \frac{1}{R},
\end{equation}
which means that the halo mass function diverges as $M \rightarrow 0 $ and does not give a cut-off as required.

Solutions by \cite{Marsh_silk_14, Bozek15, Du_newpaper} involve using a mass dependent critical density to suppress the halo mass function at lower masses. The new critical density is defined as
\begin{equation}
\delta_c^{fdm} (M,z) = G(k,z) \delta_c^{cdm}(z),
\end{equation}
where $G(k,z)$ is given as
\begin{equation}
G(k,z) = \frac{\delta_{cdm}(k,z)\delta_{cdm}(k_0,z_h)}{\delta_{cdm}(k,z_h)\delta_{cdm}(k_0,z)}/ \frac{\delta_{fdm}(k,z)\delta_{fdm}(k_0,z_h)}{\delta_{fdm}(k,z_h)\delta_{fdm}(k_0,z)},
\end{equation}
where where $k_0 = \SI{0.002}{h/Mpc}$ is a pivot scale, and $z_h$ is chosen to be
large enough so that at the relevant redshift the shape of CDM power spectrum has frozen in, selected to be 300 by \cite{Du_newpaper}.
The argument used for using this critical density is as follows: For CDM, we can take $\delta_c = 1.686$ to be a constant and take $\sigma(R,z) \propto D_{+}(z)$, or we can make $\delta_c$ redshift dependent as $\delta_c \propto 1/D_{+}(z)$ and $\sigma(R)$ to be redshift independent. The growth rate for CDM is scale independent and hence we can use a scale-independent critical density. On the other hand, FDM growth is scale-dependent and hence one should use a critical density which is higher for lower masses. 

\cite{Marsh_silk_14} and \cite{Bozek15} use this mass dependent critical density and fitting function $f(\sigma)$ given by \cite{Sheth_tormen2002}. \cite{Du_newpaper} use the same mass dependent critical density, but argue that the same fitting function cannot be used. The fitting function $f(\sigma)$ and the critical density (barrier) are related to each other in the Press-Schechter formalism as:

\begin{equation}
\int_0^S f(S') dS' + \int_{-\infty}^{B(S)} P(\delta, S) d\delta = 1,
\end{equation}
where $S = \sigma^2$, $B(S)$ is the mass dependent barrier and $P(\delta, S)$ is the probability for a trajectory to lie between $\delta$ and $\delta + d\delta$ for variance $S$. \cite{Du_newpaper} numerically calculate $f(\sigma)$ for the mass dependent critical density. This changes few properties of the halo mass function.

Both calculations by \cite{Marsh_silk_14, Bozek15} and \cite{Du_newpaper} give sharp cut-offs in the halo mass function and the cut-offs increase as we go to higher redshifts. The cut-off for \cite{Marsh_silk_14} and \cite{Bozek15} is $\sim \SI{2e8}{M_{\odot}}$ for $m_a = \SI{e-22}{eV}$ at $z = 0$. Whereas one calculated by \cite{Du_newpaper} is about four times higher. For redshift 14, \cite{Marsh_silk_14} and \cite{Bozek15} obtain a cut-off at $\SI{2e9}{M_{\odot}}$, whereas \cite{Du_newpaper} have it to be $\SI{3e9}{M_{\odot}}$.

\cite{Du_newpaper} calculate the cut-offs to be higher than those calculated by \cite{Marsh_silk_14} or \cite{Bozek15}. It is also worth noting that the cut-off mass changes less strongly with redshift in \cite{Du_newpaper}.

\subsection{Shortcomings of previous calculations}

The argument for a mass dependent critical density is based on the fact that $\sigma(R,z) \propto D_{+}(z)$ or inversely for $\delta_c$ for CDM and that this growth factor $D_{+}(z)$ should be replaced with $D_{+}(k,z)$ for FDM. Let us take a closer look at this argument.

For fuzzy dark matter, $\sigma(R)$ becomes nearly constant for $R$ smaller than the cut-off scale in the power spectrum. The shape of $\sigma(R)$ is very weakly dependent on the shape of the power spectrum for scales smaller than the cut-off scale, as the spherical top hat window draws contributions mainly from large scales.
\begin{align}
\sigma(R) \approx \sigma(R_0) & \quad \textrm{for} \, R < R_0
\end{align}

Hence, the redshift dependence of $\sigma(R)$ can be given as
\begin{equation}
\sigma(R,z) = 
\left\{
\begin{split}
\frac{D_{+}(R,z)}{D_{+}(R,z_0)} \sigma(R,z_0) && \textrm{for } R > R_0, \\
\frac{D_{+}(R_0,z)}{D_{+}(R_0,z_0)} \sigma(R,z_0) && \textrm{for } R < R_0.\\
\end{split}
\right.
\end{equation}

\begin{equation}
\therefore \sigma(R,z) = \frac{D_{+}(z)}{D_{+}(z_0)} \sigma(R,z_0).
\end{equation}

Here, $R_0$ corresponds to the scale where the FDM power spectrum starts to differ from CDM (approximately the Jeans scale at the matter-radiation equality). The scales larger than or equal to $R_0$ grow the same as CDM, which concludes that $\sigma(R)$ for FDM grows in a fashion similar to CDM with redshift. Hence, the critical density also cannot be dependent on the scale through the FDM growth rate.

It is also interesting to note that the redshift dependence in the halo mass functions calculated by \cite{Marsh_silk_14, Du_newpaper} comes mainly from the redshift dependence of the Jeans scale. The shape of the power spectrum and $\sigma(R)$ does not change with significantly the redshift, as the transfer function is nearly redshift independent \citep{Hu2000_fdm}. However, the shape of the critical density barrier changes with redshift, as it is based on which scales are growing and which are not. In their Figure 3, \cite{Du_newpaper} show the halo mass function calculated using a redshift dependent transfer function, whereas in Figure 5, they use the transfer function given by \cite{Hu2000_fdm}. For both the cases, for $m_a = \SI{e-22}{eV}$ at $z = 0$, they find a cutoff in the halo mass function at $6 \times 10^8 h^{-1} M_{\odot}$. Therefore, we conclude that the redshift dependence in the cutoff of the halo mass function in \cite{Du_newpaper} arises primarily not from the redshift evolution of the transfer function, but from the redshift evolution of the Jeans mass and the mass-dependent barrier.

In recent years, there have been many simulations that solve the Schr{\"o}dinger-Poisson equations for accurately evolving FDM \citep{Schive_nature, Schive14_soliton, Mocz17,XinyuLi19}. Although they accurately evolve fuzzy dark matter and reproduce the soliton profiles in the halos, they do not have a sufficient number of halos to calculate the halo mass function because of their small box sizes. A number of works \citep{Schive16_simulation, Sarkar16, Zhang18, Nori19} have run collision-less N-body simulations with the initial power spectrum of FDM. These numerical simulations predict many lower mass halos too. \cite{Schive16_simulation} mark and delete those `spurious' halos and calculate the halo mass function for `genuine' halos. Hence, we use the halo mass function estimated in \cite{Schive16_simulation} to compare with our results.

\cite{Du_newpaper} return their halo mass function for $m_{a} = \SI{e-22}{eV}$, whereas \cite{Schive16_simulation} return their results for $m_{a} = 0.8, 1.6, 3 \times \SI{e-22}{eV}$. We compare the results for $m_{a} = \SI{0.8e-22}{eV}$ with results from \cite{Du_newpaper}. \cite{Du_newpaper} calculate the cut-off mass for redshift $4$ to be $\sim \SI{1.5e9}{M_{\odot}}$. This cut-off will be slightly higher for $m_{a} = \SI{0.8e-22}{eV}$. \cite{Schive16_simulation} have halos as small as $\SI{4e8}{M_{\odot}}$ which is clearly smaller than the predicted cut-off from \cite{Du_newpaper}. 

\subsection{HMF using a sharp-k window function}

Instead of using the spherical top hat function and a mass dependent critical density, we use a sharp-k window function and a constant critical density in this work. This approach has been previously used for the warm dark matter \citep{Benson13, Schneider13} and was also used for the fuzzy dark matter recently by \cite{Linares21}.

\begin{equation}
W_R(k) = 
\left\{
\begin{split}
1 && \textrm{for } k \leq k_0\\
0 && \textrm{for } k > k_0\\
\end{split}
\right.
\end{equation}

\begin{figure}
\includegraphics[width=\linewidth]{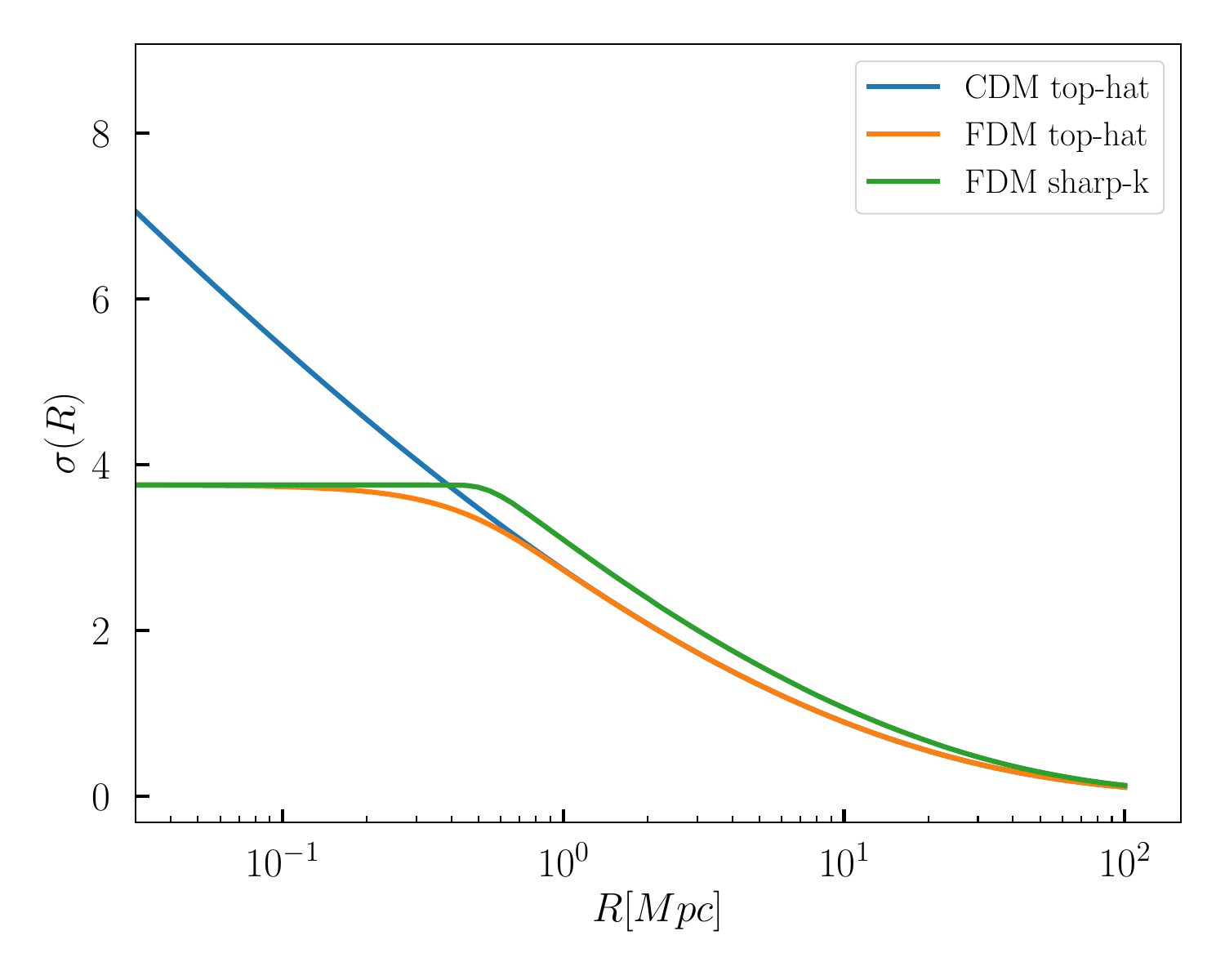}
\caption{$\sigma(R)$ using top hat window function for CDM and FDM as well as $\sigma(R)$ for FDM using sharp-k window function.}
\label{fig:sigmas}
\end{figure}

A drawback of the sharp-k window is that the enclosed mass is not clearly defined in it. It has contributions from all scales in real space, making it difficult to assign a mass $M$ for a given $k_0$. A similar problem is faced by a Gaussian window function, but it can be assigned a mass based on its integration in real space. The integration of the sharp-k window function in real space diverges \citep{Maggiore2010}. Previous works \citep{Benson13, Schneider13} using a sharp-k window with warm dark matter use $k_0 = \alpha/R$ keeping a free parameter to be fit by the numerical simulations. Following \cite{Benson13}, we use $\alpha = 2.5$, which is also close to the value of 2.42 suggested by \cite{Lacey_Cole94} using an approximate integral. We also need to rescale the critical density, as $\sigma(R)$ with sharp-k window is higher than that of top hat window for higher masses (see Figure \ref{fig:sigmas}). We rescale $\delta_c$ by multiplying it with 1.195. This is also to ensure that the halo mass function matches with CDM at higher masses.

\begin{figure}
\begin{center}
\includegraphics[width=\linewidth]{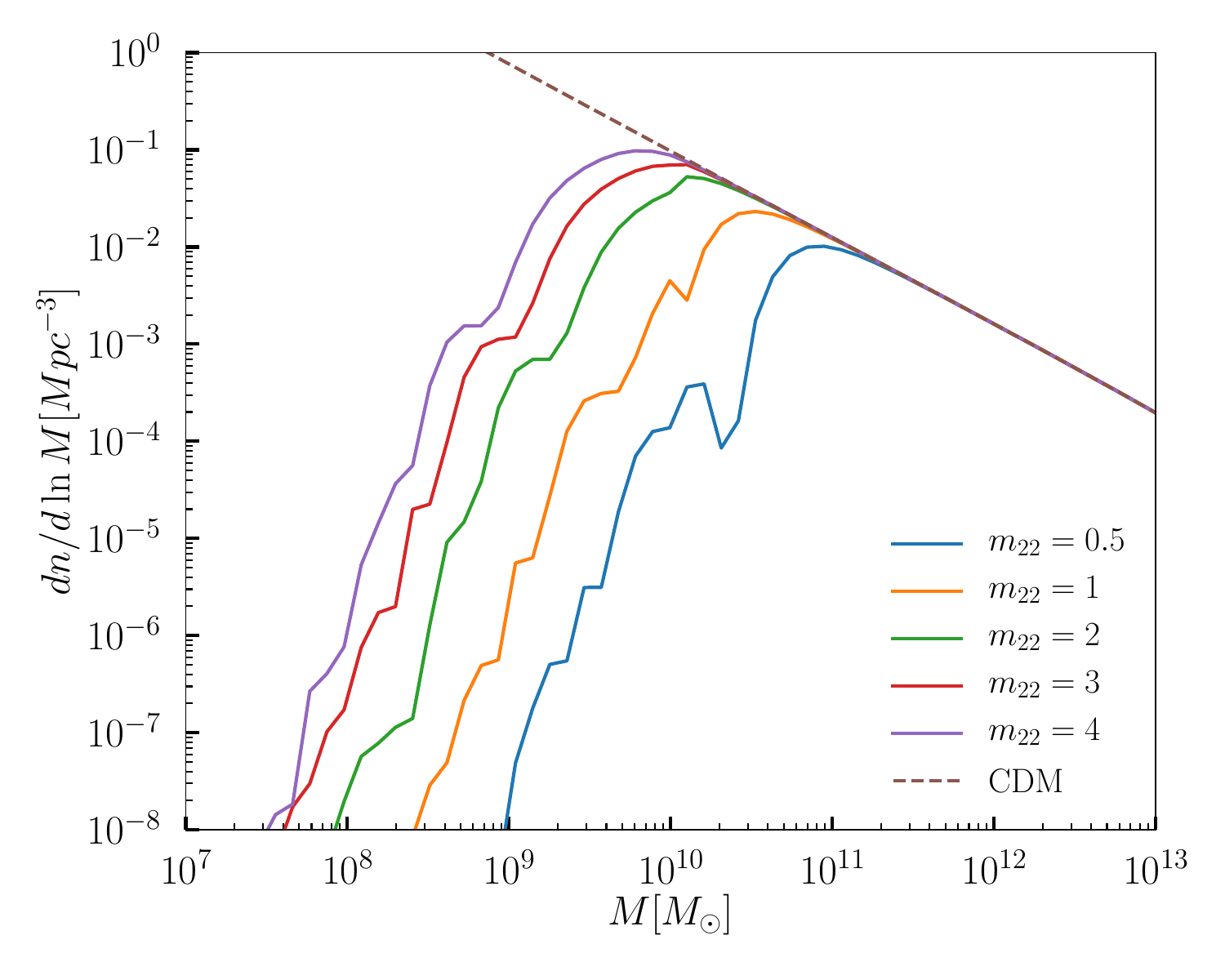}
\caption{Halo mass functions calculated using the sharp-k window function for $m_{a}$ of 0.5, 1, 2, 3, 4 $\times 10^{-22} \si{eV}$ for FDM and for CDM at $z = 0$.}
\label{fig:hmf_diff_m}
\end{center}
\end{figure}

\begin{figure}
\begin{center}
\includegraphics[width=\linewidth]{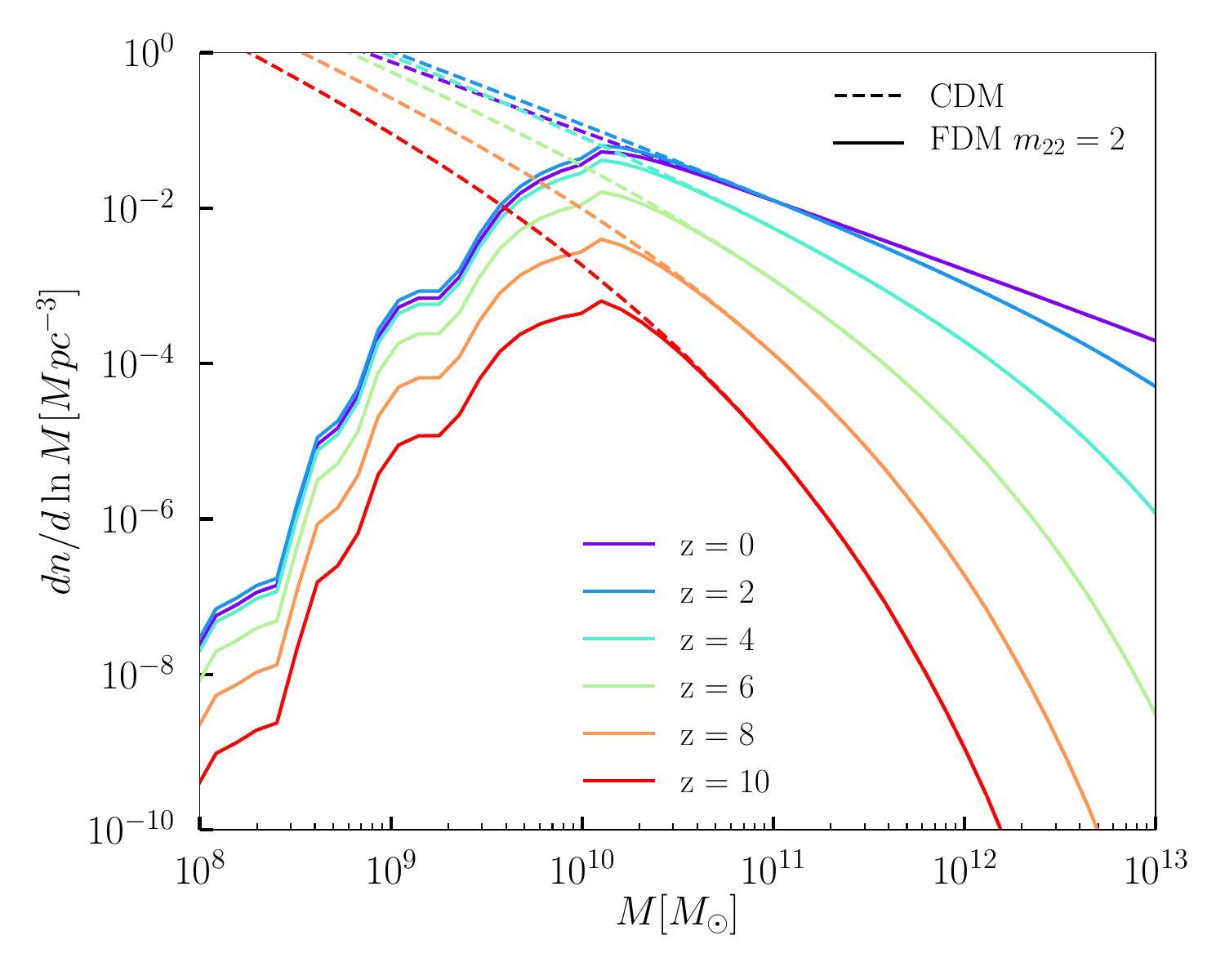}
\caption{Halo mass functions calculated using the sharp-k window function for various redshifts for $m_a = \SI{2e-22}{eV}$. The halo mass function is smoothed for masses lower than $5\times10^{10} M_{\odot}$}.
\label{fig:hmf_diff_z}
\end{center}
\end{figure}

This halo mass function does not give a sharp cut-off like \cite{Marsh_silk_14, Bozek15, Du_newpaper}. Figure \ref{fig:hmf_diff_m} shows the halo mass function for different FDM mass. The cut-off does not depend on redshift as strongly as previous works. Figure \ref{fig:hmf_diff_z} shows FDM halo mass functions for different redshifts for $m_{a} = \SI{2e-22}{eV}$.

\begin{figure}
\begin{center}
\includegraphics[width=\linewidth]{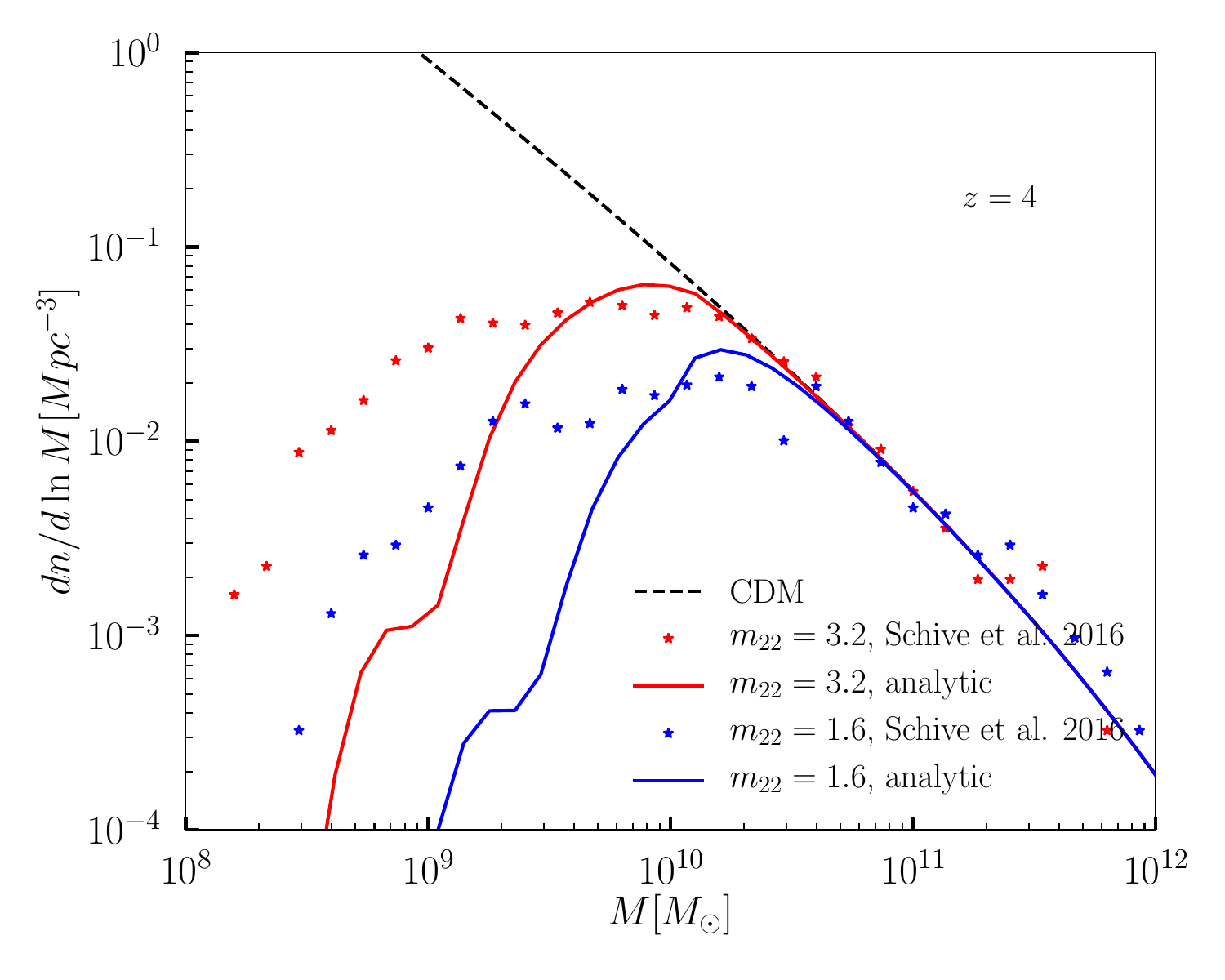}
\caption{Halo mass functions for $m_{a} = \SI{1.6e-22}{eV}$ and $m_{a} = \SI{3.2e-22}{eV}$ at redshift 4 from analytic calculations using a sharp-k window function, and from simulations by \protect\cite{Schive16_simulation}.}
\label{fig:compare_sim}
\end{center}
\end{figure}

The calculated halo mass function is shown in figure  \ref{fig:compare_sim} along with the halo mass function calculated by \cite{Schive16_simulation} for $m_{a} = \SI{3.2e-22}{eV}$ and $m_{a} = \SI{1.6e-22}{eV}$ at $z = 4$. \cite{Schive16_simulation} use collision-less dark matter simulation with an initial power spectrum for fuzzy dark matter.  

We note that the analytical halo mass function we have derived predicts a lower number of halos compared to the numerical simulations at low masses as shown in Figure \ref{fig:compare_sim}. There could be two possible reasons for this: 
\begin{itemize}
\item In their simulation, \cite{Schive16_simulation} find many low mass halos. They flag ``spurious'' halos and remove them from their calculations. It is possible that a sufficient number of halos have not been removed. A full non-linear simulation solving the Schr{\"o}dinger-Poisson equations is required to make an accurate comparison.
\item Fuzzy dark matter power spectrum falls much more rapidly as compared to the warm dark matter. Hence, halo formation from non-linear effects such as fragmentation of larger structures could be significant, and this is not considered in this calculation. This could increase the number of halos at low masses. This needs further investigation.
\end{itemize}

We agree with the shortcomings of our model to calculate the halo mass function using a sharp-k window function, particularly below the turnover halo mass scale. It needs a further investigation of the non-linear structure formation and an accurate estimation of the halo mass function from the simulations of fuzzy dark matter that use Schr{\"o}dinger-Poisson equations. Recent work by \cite{May_Springel21} suggests that the turnover in the halo mass function at the low mass end may not be as steep as we find here and finds a higher number of low mass halos. For this exercise however, we will use the analytical halo mass function that we have calculated to constrain the mass of FDM.

\section{Limits on the FDM mass}

The fuzzy dark matter paradigm has only one parameter $m_{a}$. The FDM particle is lighter than $\SI{e-18}{eV}$ and massive than $\SI{e-33}{eV}$ \citep{Marsh_review}. The mass can further be constrained by various astrophysical processes including the following. \cite{Schive16_simulation, Bozek15} use galaxy UV luminosity function and \cite{Schive16_simulation} constrain it to be $>\SI{1.2e-22}{eV}$. \cite{Sarkar16} found $m_{a}>\SI{e-23}{eV}$ using damped Lyman-$\alpha$ observations and simulations. \cite{Calabrese16} have used the cored density profile from \cite{Schive14_soliton} and observations of newly discovered ultra faint dSphs and estimated $m_{a}$ to be $3.7-5.6 \times \SI{e-22}{eV}$. \cite{Amorisco18} constrain $m_a > \SI{1.5e-22}{eV}$ based on the dynamics of stellar streams in the Milky Way. Observations of high-resolution Lyman-$\alpha$ spectra were used to  constrain the axion mass to be more than $\SI{e-21}{eV}$ \citep{Armengaud17, Irsic17, Kobayashi17}, although further work needs to be done to understand the effects of baryonic physics better. In particular, late reionization can leave a patchy distribution of neutral hydrogen that can be falsely interpreted as due to gravitational fluctuations.

In this work, we constrain the mass of axions using observations of high-redshift lensed galaxies following a method used by \cite{Lensing_pacucci13} to constrain the warm dark matter.

Cluster Lensing And Supernova survey with Hubble (CLASH) is a survey using the Hubble telescope. \cite{Zheng12} report the observation of a galaxy at redshift 9.6 with a magnification of 15 (MACS 1149-JD). \cite{Coe13} report the observation of another galaxy at redshift $\sim 10.8$ with a magnification of 8 (MACS0647-JD). \cite{Coe13} estimate intrinsic (unlensed) magnitude of MACS0647-JD to be 28.2 in F160W band and calculate the rest frame UV luminosity to be $L_{UV} \sim 2.8 \times 10^{28} \textrm{erg}\, \textrm{s}^{-1} \textrm{Hz}^{-1}$. They conclude that the stellar mass of galaxy is most likely $10^8 - 10^9 \si{M_{\odot}}$ and expect the dark halo mass to be $\sim \SI{e10}{M_{\odot}}$. We do not use the mass of the galaxy for constraining the axion mass. The magnification factor can be used to calculate the effective volume of the galaxies. Existence of galaxies at $z \sim 10$ with their effective volume can be used to constrain the fuzzy dark matter, even without explicit knowledge about masses of the galaxies. 

\begin{center}
\begin{tabular}{c c c}
Object ID & $\mu$ & $V_{eff}$ (Mpc$^3$)\\
\hline
MACS1149-JD & 15 & $\sim 700$\\
MACS0647-JD & 8 & $\sim 2000$
\end{tabular}
\end{center}

\begin{equation}
n(m_{a},z) = \int_0^{\infty} \frac{dn}{d\ln M}(M,z,m_{a})\, d\ln M
\end{equation}

We calculate the halo mass function for the fuzzy dark matter at redshift $10$. We integrate the halo mass function to get the integrated number density of halos as a function of axion mass. Since this includes halos of all masses, the number density calculated from observations cannot be higher than this.

\begin{figure}
\includegraphics[width=\linewidth]{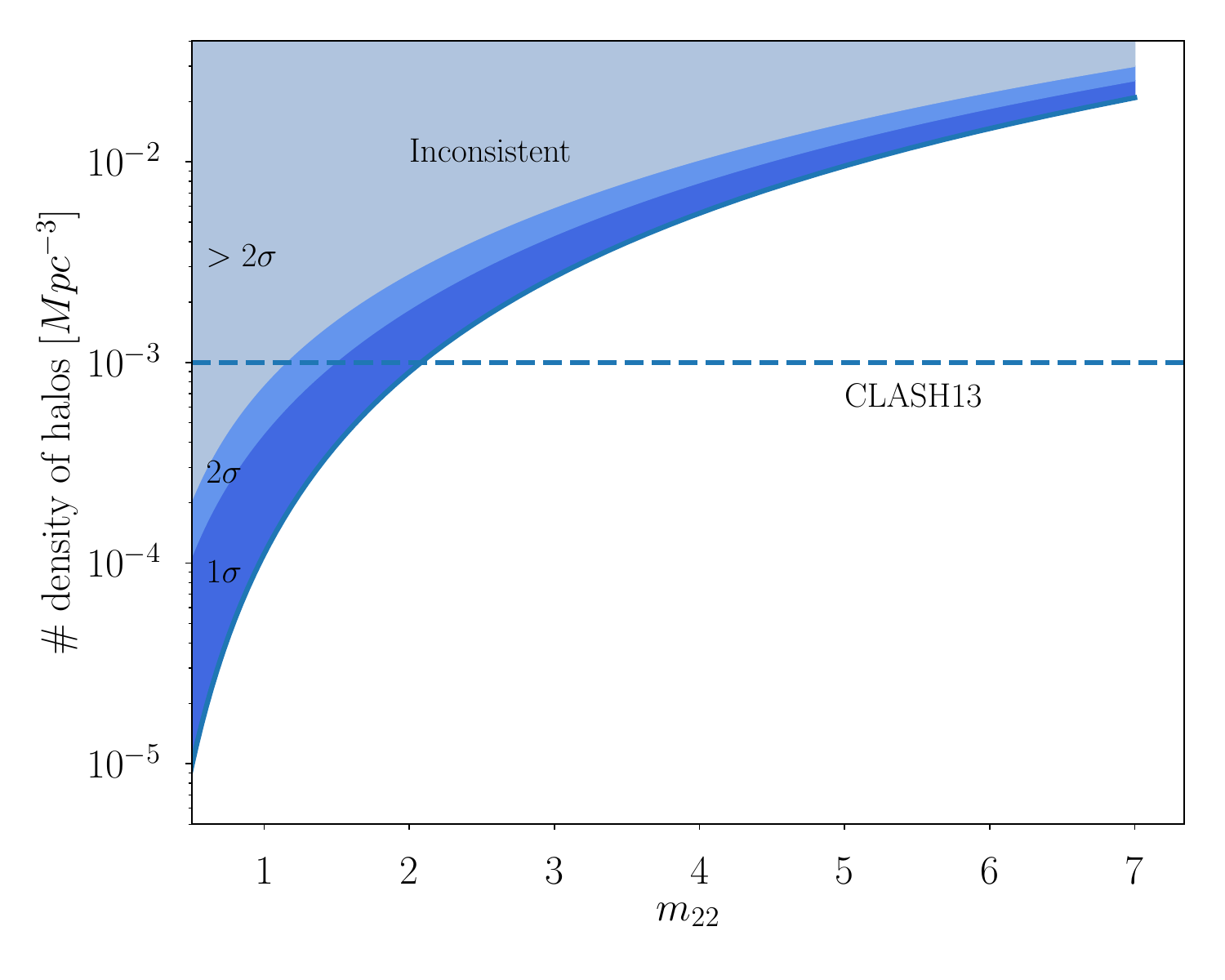}
\caption{Integrated number density of halos as a function of axion mass at $z = 10$. The horizontal line at $\SI{e-3}{Mpc^{-3}}$ denotes the number density from observations. The errors are Poisson errors.}
\label{fig:cum_dens}
\end{figure}

If we conservatively use the effective volume for two galaxies to be $\SI{2000}{Mpc^3}$, that gives $n_{tot} = \SI{2/2000}{{Mpc}^{-3}} = \SI{e-3}{{Mpc}^{-3}}$. Figure \ref{fig:cum_dens} shows a plot of cumulative number density as a function of FDM mass. We calculate Poisson errors on the number of halos in the corresponding volume i.e. $\SI{1000}{{Mpc}^3}$.  The intersection of the integrated number density and horizontal line corresponding to CLASH survey gives a lower limit on the axion mass. With this exercise we calculate to be $m_a > \SI{2e-22}{eV}$. With $2-\sigma$ error (95\% confidence interval), the minimum mass of axion is constrained to be $\SI{1.2e-22}{eV}$. 

While calculating this constraint, we have not used any astrophysical processes other than the halo mass function. We have integrated over the entire halo mass function. Tighter constraints can be obtained by accurately simulating the stellar properties of first galaxies in FDM, observations of lensed galaxies at higher redshifts and geometries of the effective volume. As the early galaxy formation is significantly delayed in a fuzzy dark matter cosmology, observations of high-redshift galaxies coupled with accurate numerical predictions may be the best way to constrain FDM. The \textit{Roman Space Telescope (WFIRST)} High Latitude Survey is expected to observe many $z \sim 10$ galaxies \citep{WFIRST13, Waters16} which can be studied in detail with the \textit{James Webb Space Telscope (JWST)}. The \textit{JWST} is also expected to measure the faint-end of the UV luminosity function at high-redshift which can also be used to constrain the properties of the fuzzy dark matter \citep{Corasaniti17, Ni19}.

\section*{Acknowledgements}

We thank Zolt{\'a}n Haiman for suggesting the method to constrain the axion mass; Lam Hui, Hsi-Yu Schive, and Tom Broadhurst for helpful discussions and Hsi-Yu Schive for providing the simulation data. 

\section*{Data availability}
No new data were generated or analysed in support of this research. Numerical code to calculate the halo mass function will be shared on reasonable request to the corresponding author.

%%%%%%%%%%%%%%%%%%%%%%%%%%%%%%%%%%%%%%%%%%%%%%%%%%



\bibliographystyle{mnras}
\bibliography{report_bib} % if your bibtex file is called example.bib

\begin{thebibliography}{}
\makeatletter
\relax
\def\mn@urlcharsother{\let\do\@makeother \do\$\do\&\do\#\do\^\do\_\do\%\do\~}
\def\mn@doi{\begingroup\mn@urlcharsother \@ifnextchar [ {\mn@doi@}
  {\mn@doi@[]}}
\def\mn@doi@[#1]#2{\def\@tempa{#1}\ifx\@tempa\@empty \href
  {http://dx.doi.org/#2} {doi:#2}\else \href {http://dx.doi.org/#2} {#1}\fi
  \endgroup}
\def\mn@eprint#1#2{\mn@eprint@#1:#2::\@nil}
\def\mn@eprint@arXiv#1{\href {http://arxiv.org/abs/#1} {{\tt arXiv:#1}}}
\def\mn@eprint@dblp#1{\href {http://dblp.uni-trier.de/rec/bibtex/#1.xml}
  {dblp:#1}}
\def\mn@eprint@#1:#2:#3:#4\@nil{\def\@tempa {#1}\def\@tempb {#2}\def\@tempc
  {#3}\ifx \@tempc \@empty \let \@tempc \@tempb \let \@tempb \@tempa \fi \ifx
  \@tempb \@empty \def\@tempb {arXiv}\fi \@ifundefined
  {mn@eprint@\@tempb}{\@tempb:\@tempc}{\expandafter \expandafter \csname
  mn@eprint@\@tempb\endcsname \expandafter{\@tempc}}}

\bibitem[\protect\citeauthoryear{{Amorisco} \& {Loeb}}{{Amorisco} \&
  {Loeb}}{2018}]{Amorisco18}
{Amorisco} N.~C.,  {Loeb} A.,  2018, arXiv e-prints, \href
  {https://ui.adsabs.harvard.edu/abs/2018arXiv180800464A} {p. arXiv:1808.00464}

\bibitem[\protect\citeauthoryear{{Armengaud}, {Palanque-Delabrouille},
  {Y{\`e}che}, {Marsh}  \& {Baur}}{{Armengaud} et~al.}{2017}]{Armengaud17}
{Armengaud} E.,  {Palanque-Delabrouille} N.,  {Y{\`e}che} C.,  {Marsh} D.
  J.~E.,   {Baur} J.,  2017, \mn@doi [\mnras] {10.1093/mnras/stx1870}, \href
  {https://ui.adsabs.harvard.edu/abs/2017MNRAS.471.4606A} {471, 4606}

\bibitem[\protect\citeauthoryear{{Bennett} et~al.,}{{Bennett}
  et~al.}{2013}]{Bennett13}
{Bennett} C.~L.,  et~al., 2013, \mn@doi [\apjs] {10.1088/0067-0049/208/2/20},
  \href {http://adsabs.harvard.edu/abs/2013ApJS..208...20B} {208, 20}

\bibitem[\protect\citeauthoryear{{Benson} et~al.,}{{Benson}
  et~al.}{2013}]{Benson13}
{Benson} A.~J.,  et~al., 2013, \mn@doi [\mnras] {10.1093/mnras/sts159}, \href
  {http://adsabs.harvard.edu/abs/2013MNRAS.428.1774B} {428, 1774}

\bibitem[\protect\citeauthoryear{{Bond}, {Cole}, {Efstathiou}  \&
  {Kaiser}}{{Bond} et~al.}{1991}]{Bond91}
{Bond} J.~R.,  {Cole} S.,  {Efstathiou} G.,   {Kaiser} N.,  1991, \mn@doi
  [\apj] {10.1086/170520}, \href
  {http://adsabs.harvard.edu/abs/1991ApJ...379..440B} {379, 440}

\bibitem[\protect\citeauthoryear{{Boylan-Kolchin}, {Bullock}  \&
  {Kaplinghat}}{{Boylan-Kolchin} et~al.}{2011}]{Boylan11_toobigtofail}
{Boylan-Kolchin} M.,  {Bullock} J.~S.,   {Kaplinghat} M.,  2011, \mn@doi
  [\mnras] {10.1111/j.1745-3933.2011.01074.x}, \href
  {http://adsabs.harvard.edu/abs/2011MNRAS.415L..40B} {415, L40}

\bibitem[\protect\citeauthoryear{{Bozek}, {Marsh}, {Silk}  \& {Wyse}}{{Bozek}
  et~al.}{2015}]{Bozek15}
{Bozek} B.,  {Marsh} D.~J.~E.,  {Silk} J.,   {Wyse} R.~F.~G.,  2015, \mn@doi
  [\mnras] {10.1093/mnras/stv624}, \href
  {http://adsabs.harvard.edu/abs/2015MNRAS.450..209B} {450, 209}

\bibitem[\protect\citeauthoryear{{Burkert}}{{Burkert}}{1995}]{Burkert95}
{Burkert} A.,  1995, \mn@doi [\apjl] {10.1086/309560}, \href
  {http://adsabs.harvard.edu/abs/1995ApJ...447L..25B} {447, L25}

\bibitem[\protect\citeauthoryear{{Calabrese} \& {Spergel}}{{Calabrese} \&
  {Spergel}}{2016}]{Calabrese16}
{Calabrese} E.,  {Spergel} D.~N.,  2016, \mn@doi [\mnras]
  {10.1093/mnras/stw1256}, \href
  {http://adsabs.harvard.edu/abs/2016MNRAS.460.4397C} {460, 4397}

\bibitem[\protect\citeauthoryear{{Coe} et~al.,}{{Coe} et~al.}{2013}]{Coe13}
{Coe} D.,  et~al., 2013, \mn@doi [\apj] {10.1088/0004-637X/762/1/32}, \href
  {http://adsabs.harvard.edu/abs/2013ApJ...762...32C} {762, 32}

\bibitem[\protect\citeauthoryear{{Corasaniti}, {Agarwal}, {Marsh}  \&
  {Das}}{{Corasaniti} et~al.}{2017}]{Corasaniti17}
{Corasaniti} P.~S.,  {Agarwal} S.,  {Marsh} D.~J.~E.,   {Das} S.,  2017,
  \mn@doi [\prd] {10.1103/PhysRevD.95.083512}, \href
  {https://ui.adsabs.harvard.edu/abs/2017PhRvD..95h3512C} {95, 083512}

\bibitem[\protect\citeauthoryear{{Du}, {Behrens}  \& {Niemeyer}}{{Du}
  et~al.}{2017}]{Du_newpaper}
{Du} X.,  {Behrens} C.,   {Niemeyer} J.~C.,  2017, \mn@doi [\mnras]
  {10.1093/mnras/stw2724}, \href
  {https://ui.adsabs.harvard.edu/abs/2017MNRAS.465..941D} {465, 941}

\bibitem[\protect\citeauthoryear{{Goerdt}, {Moore}, {Read}, {Stadel}  \&
  {Zemp}}{{Goerdt} et~al.}{2006}]{Goerdt2006}
{Goerdt} T.,  {Moore} B.,  {Read} J.~I.,  {Stadel} J.,   {Zemp} M.,  2006,
  \mn@doi [\mnras] {10.1111/j.1365-2966.2006.10182.x}, \href
  {http://adsabs.harvard.edu/abs/2006MNRAS.368.1073G} {368, 1073}

\bibitem[\protect\citeauthoryear{{Hinshaw} et~al.,}{{Hinshaw}
  et~al.}{2013}]{WMAP9}
{Hinshaw} G.,  et~al., 2013, \mn@doi [\apjs] {10.1088/0067-0049/208/2/19},
  \href {http://adsabs.harvard.edu/abs/2013ApJS..208...19H} {208, 19}

\bibitem[\protect\citeauthoryear{{Hu}, {Barkana}  \& {Gruzinov}}{{Hu}
  et~al.}{2000}]{Hu2000_fdm}
{Hu} W.,  {Barkana} R.,   {Gruzinov} A.,  2000, \mn@doi [Physical Review
  Letters] {10.1103/PhysRevLett.85.1158}, \href
  {http://adsabs.harvard.edu/abs/2000PhRvL..85.1158H} {85, 1158}

\bibitem[\protect\citeauthoryear{{Hui}, {Ostriker}, {Tremaine}  \&
  {Witten}}{{Hui} et~al.}{2017}]{Hui17}
{Hui} L.,  {Ostriker} J.~P.,  {Tremaine} S.,   {Witten} E.,  2017, \mn@doi
  [\prd] {10.1103/PhysRevD.95.043541}, \href
  {https://ui.adsabs.harvard.edu/abs/2017PhRvD..95d3541H} {95, 043541}

\bibitem[\protect\citeauthoryear{{Ir{\v{s}}i{\v{c}}}, {Viel}, {Haehnelt},
  {Bolton}  \& {Becker}}{{Ir{\v{s}}i{\v{c}}} et~al.}{2017}]{Irsic17}
{Ir{\v{s}}i{\v{c}}} V.,  {Viel} M.,  {Haehnelt} M.~G.,  {Bolton} J.~S.,
  {Becker} G.~D.,  2017, \mn@doi [\prl] {10.1103/PhysRevLett.119.031302}, \href
  {https://ui.adsabs.harvard.edu/abs/2017PhRvL.119c1302I} {119, 031302}

\bibitem[\protect\citeauthoryear{{Khlopov}, {Malomed}  \&
  {Zeldovich}}{{Khlopov} et~al.}{1985}]{Khlopov85}
{Khlopov} M.~I.,  {Malomed} B.~A.,   {Zeldovich} I.~B.,  1985, \mn@doi [\mnras]
  {10.1093/mnras/215.4.575}, \href
  {https://ui.adsabs.harvard.edu/abs/1985MNRAS.215..575K} {215, 575}

\bibitem[\protect\citeauthoryear{{Klypin}, {Kravtsov}, {Valenzuela}  \&
  {Prada}}{{Klypin} et~al.}{1999}]{Klypin99}
{Klypin} A.,  {Kravtsov} A.~V.,  {Valenzuela} O.,   {Prada} F.,  1999, \mn@doi
  [\apj] {10.1086/307643}, \href
  {http://adsabs.harvard.edu/abs/1999ApJ...522...82K} {522, 82}

\bibitem[\protect\citeauthoryear{{Kobayashi}, {Murgia}, {De Simone},
  {Ir{\v{s}}i{\v{c}}}  \& {Viel}}{{Kobayashi} et~al.}{2017}]{Kobayashi17}
{Kobayashi} T.,  {Murgia} R.,  {De Simone} A.,  {Ir{\v{s}}i{\v{c}}} V.,
  {Viel} M.,  2017, \mn@doi [\prd] {10.1103/PhysRevD.96.123514}, \href
  {https://ui.adsabs.harvard.edu/abs/2017PhRvD..96l3514K} {96, 123514}

\bibitem[\protect\citeauthoryear{{Lacey} \& {Cole}}{{Lacey} \&
  {Cole}}{1994}]{Lacey_Cole94}
{Lacey} C.,  {Cole} S.,  1994, \mn@doi [\mnras] {10.1093/mnras/271.3.676},
  \href {https://ui.adsabs.harvard.edu/abs/1994MNRAS.271..676L} {271, 676}

\bibitem[\protect\citeauthoryear{{Li}, {Hui}  \& {Bryan}}{{Li}
  et~al.}{2019}]{XinyuLi19}
{Li} X.,  {Hui} L.,   {Bryan} G.~L.,  2019, \mn@doi [\prd]
  {10.1103/PhysRevD.99.063509}, \href
  {https://ui.adsabs.harvard.edu/abs/2019PhRvD..99f3509L} {99, 063509}

\bibitem[\protect\citeauthoryear{{Linares Cede{\~n}o}, {Gonz{\'a}lez-Morales}
  \& {Ure{\~n}a-L{\'o}pez}}{{Linares Cede{\~n}o} et~al.}{2021}]{Linares21}
{Linares Cede{\~n}o} F.~X.,  {Gonz{\'a}lez-Morales} A.~X.,
  {Ure{\~n}a-L{\'o}pez} L.~A.,  2021, \mn@doi [\jcap]
  {10.1088/1475-7516/2021/01/051}, \href
  {https://ui.adsabs.harvard.edu/abs/2021JCAP...01..051L} {2021, 051}

\bibitem[\protect\citeauthoryear{{Macci{\`o}}, {Paduroiu}, {Anderhalden},
  {Schneider}  \& {Moore}}{{Macci{\`o}} et~al.}{2012}]{Macci2012_catch22}
{Macci{\`o}} A.~V.,  {Paduroiu} S.,  {Anderhalden} D.,  {Schneider} A.,
  {Moore} B.,  2012, \mn@doi [\mnras] {10.1111/j.1365-2966.2012.21284.x}, \href
  {http://adsabs.harvard.edu/abs/2012MNRAS.424.1105M} {424, 1105}

\bibitem[\protect\citeauthoryear{{Maggiore} \& {Riotto}}{{Maggiore} \&
  {Riotto}}{2010}]{Maggiore2010}
{Maggiore} M.,  {Riotto} A.,  2010, \mn@doi [\apj]
  {10.1088/0004-637X/711/2/907}, \href
  {http://adsabs.harvard.edu/abs/2010ApJ...711..907M} {711, 907}

\bibitem[\protect\citeauthoryear{{Marsh}}{{Marsh}}{2016}]{Marsh_review}
{Marsh} D. J.~E.,  2016, \mn@doi [\physrep] {10.1016/j.physrep.2016.06.005},
  \href {https://ui.adsabs.harvard.edu/abs/2016PhR...643....1M} {643, 1}

\bibitem[\protect\citeauthoryear{{Marsh} \& {Silk}}{{Marsh} \&
  {Silk}}{2014}]{Marsh_silk_14}
{Marsh} D.~J.~E.,  {Silk} J.,  2014, \mn@doi [\mnras] {10.1093/mnras/stt2079},
  \href {http://adsabs.harvard.edu/abs/2014MNRAS.437.2652M} {437, 2652}

\bibitem[\protect\citeauthoryear{{May} \& {Springel}}{{May} \&
  {Springel}}{2021}]{May_Springel21}
{May} S.,  {Springel} V.,  2021, \mn@doi [\mnras] {10.1093/mnras/stab1764},
  \href {https://ui.adsabs.harvard.edu/abs/2021MNRAS.506.2603M} {506, 2603}

\bibitem[\protect\citeauthoryear{{Mocz}, {Vogelsberger}, {Robles}, {Zavala},
  {Boylan-Kolchin}, {Fialkov}  \& {Hernquist}}{{Mocz} et~al.}{2017}]{Mocz17}
{Mocz} P.,  {Vogelsberger} M.,  {Robles} V.~H.,  {Zavala} J.,  {Boylan-Kolchin}
  M.,  {Fialkov} A.,   {Hernquist} L.,  2017, \mn@doi [\mnras]
  {10.1093/mnras/stx1887}, \href
  {https://ui.adsabs.harvard.edu/abs/2017MNRAS.471.4559M} {471, 4559}

\bibitem[\protect\citeauthoryear{{Navarro}, {Eke}  \& {Frenk}}{{Navarro}
  et~al.}{1996}]{Navarro96}
{Navarro} J.~F.,  {Eke} V.~R.,   {Frenk} C.~S.,  1996, \mn@doi [\mnras]
  {10.1093/mnras/283.3.L72}, \href
  {http://adsabs.harvard.edu/abs/1996MNRAS.283L..72N} {283, L72}

\bibitem[\protect\citeauthoryear{{Navarro}, {Frenk}  \& {White}}{{Navarro}
  et~al.}{1997}]{NFW}
{Navarro} J.~F.,  {Frenk} C.~S.,   {White} S.~D.~M.,  1997, \apj, \href
  {http://adsabs.harvard.edu/abs/1997ApJ...490..493N} {490, 493}

\bibitem[\protect\citeauthoryear{{Ni}, {Wang}, {Feng}  \& {Di Matteo}}{{Ni}
  et~al.}{2019}]{Ni19}
{Ni} Y.,  {Wang} M.-Y.,  {Feng} Y.,   {Di Matteo} T.,  2019, \mn@doi [\mnras]
  {10.1093/mnras/stz2085}, \href
  {https://ui.adsabs.harvard.edu/abs/2019MNRAS.488.5551N} {488, 5551}

\bibitem[\protect\citeauthoryear{{Nori}, {Murgia}, {Ir{\v{s}}i{\v{c}}}, {Baldi}
   \& {Viel}}{{Nori} et~al.}{2019}]{Nori19}
{Nori} M.,  {Murgia} R.,  {Ir{\v{s}}i{\v{c}}} V.,  {Baldi} M.,   {Viel} M.,
  2019, \mn@doi [\mnras] {10.1093/mnras/sty2888}, \href
  {https://ui.adsabs.harvard.edu/abs/2019MNRAS.482.3227N} {482, 3227}

\bibitem[\protect\citeauthoryear{{Pacucci}, {Mesinger}  \& {Haiman}}{{Pacucci}
  et~al.}{2013}]{Lensing_pacucci13}
{Pacucci} F.,  {Mesinger} A.,   {Haiman} Z.,  2013, \mn@doi [\mnras]
  {10.1093/mnrasl/slt093}, \href
  {http://adsabs.harvard.edu/abs/2013MNRAS.435L..53P} {435, L53}

\bibitem[\protect\citeauthoryear{{Press} \& {Schechter}}{{Press} \&
  {Schechter}}{1974}]{Press_schechter74}
{Press} W.~H.,  {Schechter} P.,  1974, \mn@doi [\apj] {10.1086/152650}, \href
  {http://adsabs.harvard.edu/abs/1974ApJ...187..425P} {187, 425}

\bibitem[\protect\citeauthoryear{{Sarkar}, {Mondal}, {Das}, {Sethi},
  {Bharadwaj}  \& {Marsh}}{{Sarkar} et~al.}{2016}]{Sarkar16}
{Sarkar} A.,  {Mondal} R.,  {Das} S.,  {Sethi} S.~K.,  {Bharadwaj} S.,
  {Marsh} D.~J.~E.,  2016, \mn@doi [\jcap] {10.1088/1475-7516/2016/04/012},
  \href {http://adsabs.harvard.edu/abs/2016JCAP...04..012S} {4, 012}

\bibitem[\protect\citeauthoryear{{Schive}, {Chiueh}  \& {Broadhurst}}{{Schive}
  et~al.}{2014a}]{Schive_nature}
{Schive} H.-Y.,  {Chiueh} T.,   {Broadhurst} T.,  2014a, \mn@doi [Nature
  Physics] {10.1038/nphys2996}, \href
  {http://adsabs.harvard.edu/abs/2014NatPh..10..496S} {10, 496}

\bibitem[\protect\citeauthoryear{{Schive}, {Liao}, {Woo}, {Wong}, {Chiueh},
  {Broadhurst}  \& {Hwang}}{{Schive} et~al.}{2014b}]{Schive14_soliton}
{Schive} H.-Y.,  {Liao} M.-H.,  {Woo} T.-P.,  {Wong} S.-K.,  {Chiueh} T.,
  {Broadhurst} T.,   {Hwang} W.-Y.~P.,  2014b, \mn@doi [Physical Review
  Letters] {10.1103/PhysRevLett.113.261302}, \href
  {http://adsabs.harvard.edu/abs/2014PhRvL.113z1302S} {113, 261302}

\bibitem[\protect\citeauthoryear{{Schive}, {Chiueh}, {Broadhurst}  \&
  {Huang}}{{Schive} et~al.}{2016}]{Schive16_simulation}
{Schive} H.-Y.,  {Chiueh} T.,  {Broadhurst} T.,   {Huang} K.-W.,  2016, \mn@doi
  [\apj] {10.3847/0004-637X/818/1/89}, \href
  {http://adsabs.harvard.edu/abs/2016ApJ...818...89S} {818, 89}

\bibitem[\protect\citeauthoryear{{Schneider}, {Smith}  \& {Reed}}{{Schneider}
  et~al.}{2013}]{Schneider13}
{Schneider} A.,  {Smith} R.~E.,   {Reed} D.,  2013, \mn@doi [\mnras]
  {10.1093/mnras/stt829}, \href
  {http://adsabs.harvard.edu/abs/2013MNRAS.433.1573S} {433, 1573}

\bibitem[\protect\citeauthoryear{{Sheth} \& {Tormen}}{{Sheth} \&
  {Tormen}}{2002}]{Sheth_tormen2002}
{Sheth} R.~K.,  {Tormen} G.,  2002, \mn@doi [\mnras]
  {10.1046/j.1365-8711.2002.04950.x}, \href
  {http://adsabs.harvard.edu/abs/2002MNRAS.329...61S} {329, 61}

\bibitem[\protect\citeauthoryear{{Spergel} et~al.,}{{Spergel}
  et~al.}{2013}]{WFIRST13}
{Spergel} D.,  et~al., 2013, arXiv e-prints, \href
  {https://ui.adsabs.harvard.edu/abs/2013arXiv1305.5425S} {p. arXiv:1305.5425}

\bibitem[\protect\citeauthoryear{{Tremaine}}{{Tremaine}}{1976}]{Tremaine76}
{Tremaine} S.~D.,  1976, \mn@doi [\apj] {10.1086/154085}, \href
  {http://adsabs.harvard.edu/abs/1976ApJ...203..345T} {203, 345}

\bibitem[\protect\citeauthoryear{{Waters}, {Di Matteo}, {Feng}, {Wilkins}  \&
  {Croft}}{{Waters} et~al.}{2016}]{Waters16}
{Waters} D.,  {Di Matteo} T.,  {Feng} Y.,  {Wilkins} S.~M.,   {Croft} R. A.~C.,
   2016, \mn@doi [\mnras] {10.1093/mnras/stw2000}, \href
  {https://ui.adsabs.harvard.edu/abs/2016MNRAS.463.3520W} {463, 3520}

\bibitem[\protect\citeauthoryear{{Zhang}, {Kuo}, {Liu}, {Sming Tsai}, {Cheung}
  \& {Chu}}{{Zhang} et~al.}{2018}]{Zhang18}
{Zhang} J.,  {Kuo} J.-L.,  {Liu} H.,  {Sming Tsai} Y.-L.,  {Cheung} K.,   {Chu}
  M.-C.,  2018, \mn@doi [\apj] {10.3847/1538-4357/aacf3f}, \href
  {https://ui.adsabs.harvard.edu/abs/2018ApJ...863...73Z} {863, 73}

\bibitem[\protect\citeauthoryear{{Zheng} et~al.,}{{Zheng}
  et~al.}{2012}]{Zheng12}
{Zheng} W.,  et~al., 2012, \mn@doi [\nat] {10.1038/nature11446}, \href
  {http://adsabs.harvard.edu/abs/2012Natur.489..406Z} {489, 406}

\makeatother
\end{thebibliography}

%%%%%%%%%%%%%%%%%%%%%%%%%%%%%%%%%%%%%%%%%%%%%%%%%%

%%%%%%%%%%%%%%%%% APPENDICES %%%%%%%%%%%%%%%%%%%%%

%%%%%%%%%%%%%%%%%%%%%%%%%%%%%%%%%%%%%%%%%%%%%%%%%%

% Don't change these lines
\bsp	% typesetting comment
\label{lastpage}
\end{document}